\documentclass[a4paper,cite]{article}
\usepackage{latexsym}

\usepackage[latin1]{inputenc}

\usepackage{graphicx}
\usepackage{color,pst-plot}

\usepackage{amsmath}
\usepackage {amsfonts,amssymb,amstext}

\newcommand{\lbl}[1]{\label{eq:#1}}
\newcommand{ \rf}[1]{(\ref{eq:#1})}

\newcommand{\be}{\begin{equation}}
\newcommand{\ee}{\end{equation}}{
\newcommand{\bea}{\begin{eqnarray}}
\newcommand{\eea}{\end{eqnarray}}
\newcommand{\setl}{\setlength\arraycolsep{2pt}}

\newcommand{\noi}{\noindent}
\newcommand{\nn}{\nonumber}
\newcommand{\ra}{\rightarrow}

\newcommand{\cO}{{\cal O}}

\newcommand{\Ree}{\mbox{\rm Re}}

\newcommand{\MeV}{\mbox{\rm MeV}}
\newcommand{\GeV}{\mbox{\rm GeV}}

\DeclareMathAlphabet{\eusm}{U}{}{}{}
\SetMathAlphabet\eusm{normal}{U}{eus}{m}{n}
\SetMathAlphabet\eusm{bold}{U}{eus}{b}{n}
\DeclareMathAlphabet{\mathpzc}{OT1}{pzc}{m}{it}

\input epsf

\setcounter{section}{0}

\setcounter{equation}{0}
\def\theequation{\arabic{section}.\arabic{equation}}

\begin{document}

\begin{titlepage}

\begin{flushright} UB-ECM-PF 07/10 \\CPT-P34-2007 \\
\end{flushright}

\vspace*{0.2cm}

\begin{center} {\Large \bf Bounds on the Light Quark Masses: \\ [0.5
 cm]
 the scalar channel revisited. }\\[2 cm]
{{\bf Alexandre Dominguez-Clarimon }$^{a}$, 
{\bf Eduardo de Rafael }$^{b}$  {\bf and Josep~Taron }$^{a}$}\\[1cm] 
${}^a$ Departament d'Estructura i Constituents de la Mat\`{e}ria\\
Facultat de F\'{\i}sica, Universitat de Barcelona\\
Diagonal 647, E-08028 Barcelona, Spain\\[0.3cm] and\\[0.3cm] 
${}^b$ Centre  de Physique Th\'eorique~\footnote{ Unité Mixte de Recherche (UMR 6207) du CNRS et des Universités Aix-Marseille~1, Aix-Marseille~2 et sud Toulon-Var, affilié à la FRUMAN.}\\ CNRS-Luminy,
Case 907\\ F-13288 Marseille Cedex 9, France\\ 
\end{center}

\vspace*{2.0cm}

\begin{abstract}
We evaluate lower bounds on the sum of the up and down quark masses. The bounds follow from the constraints provided by the dispersion relation obeyed by the two--point function
of the scalar current density when combined with properties of the scalar spectral function at long--distances and perturbative QCD at short--distances. Our results 
point to values of $m_u+m_d$ somewhat higher than those reported recently in the literature using lattice QCD simulations. 

\end{abstract}

\vfill

\begin{flushleft} \today\\
\end{flushleft}

\end{titlepage}

\section{\normalsize Introduction}\lbl{int}

The light $u$,$d$ and $s$ quark  masses are fundamental parameters
of the Standard Model of Particle Physics whose precise values are still
affected by large uncertainties. 
The difficulty arises from the fact that it requires 
to have a good handle on QCD non--perturbative effects in order to extract
their values
(much smaller than the typical hadronic
scale of a few hundred MeV) from the properties of the hadrons in which they
are confined.
The masses we refer to are the ones in the QCD Lagrangian and it is 
customary to
report their values in a mass--independent subtraction scheme such as the
$\overline{{\rm MS}}$--scheme at a reference scale $\mu$  which has conventionally been fixed by the lattice community at $\mu =2~\GeV$, and which we also adopt.

The extraction of the light quark masses from physical observables has been attempted 
using three independent methods: spectral function sum rules for hadronic 
correlation functions, lattice simulations, and  chiral perturbation theory
(ChPT). A summary of the spread of 
results obtained (also within each method) can be found in the successive versions of the Particle Data Book~\cite{ParticleDataBook} with references therein. The suggestion that ChPT can be used to extract ratios of the light quark masses from the physical properties of the low pseudoscalar particles goes back to earlier work by Weinberg~\cite{Wein} which has been subsequently improved (see e.g. ref.~\cite{Leut} and references therein). ChPT can thus relate determinations of $m_u +m_d$ to those of $m_u +m_s$ in a reliable way. 

We shall be concerned with the question: {\it how small can the light quark masses be?} This question was already addressed in an earlier paper~\cite{LdeRT97} where a variety of lower bounds were derived. Here we propose to reexamine the bounds concerning the scalar--isoscalar channel where a number of new developments have appeared in the meantime. On the one hand, from a phenomenological point of view, the scalar pion form factor which governs the dominant two--pion contribution to the spectral function in the scalar two--point function, has been the object of careful analyses (see section III below). On the other hand, from a theoretical point of view, the two--point function in question has been evaluated at the five loop level~\cite{CHETYRKIN,CHETYRKIN2} in perturbative QCD (pQCD), thus providing a better handle on the choice of the euclidean $Q^2$--value at which one can be confident that the perturbative regime applies.

\section{\normalsize Bounds from the Scalar Channel}\lbl{BSC}
\setcounter{equation}{0}
\def\theequation{\arabic{section}.\arabic{equation}}

The scalar density current operator in question is
\begin{equation}\lbl{eq:current}
S(x)=\hat{m} \left[ \bar{u} u + \bar{d} d \right](x),
\;\;\; \hat{m}\equiv \frac{m_u+m_d}{2}\,,
\end{equation}
with
\begin{equation}\label{dospunts}
\Psi(q^2)= i  \int  d^4x \; e^{iq \cdot x} 
\langle 0 \vert T(S(x) S(0)) |0 \rangle\,,
\end{equation}
the associated two-point function.
This function obeys a dispersion relation which in QCD requires
two subtractions. In terms of its second derivative, which gets rid of the two subtractions, and for Euclidean
values of $Q^2 \equiv - q^2$, the dispersion relation then reads
\begin{equation}\lbl{eq:dispersio}
\Psi''(Q^2)\equiv \left(\frac{\partial ^2}{(\partial q^2)^2}\Ree\Psi(q^2)\right)_{q^2=-Q^2}=\int_{(2 m_\pi)^2}^{\infty}  dt  
\frac{2}{(t+Q^2)^3} \; \frac{1}{\pi} {\rm Im}
\Psi(t)\,.
\end{equation}
This is the master equation from which the bounds on $\hat{m}$ will
be extracted.

The behaviour of the $\Psi''(Q^2)$ function in pQCD at large euclidean values: $Q^2 >> \Lambda_{\rm QCD}^2$, is known from more and more refined
calculations (see refs.~\cite{JEFE3} to~\cite{26} and ~\cite{CHETYRKIN,CHETYRKIN2}) 
\begin{equation}\lbl{psipp}
\Psi''(Q^2)= \frac{N_c}{4 \pi^2} \frac{\hat{m}^2(Q^2)}{Q^2}
\left[ \; 1 + \frac{11}{3} \frac{\alpha_s(Q^2)}{\pi} + \cdots \right]\,,
\end{equation}
where the dots represent  higher order terms which are now known to
${\cal O}(\alpha_s^4)$.
Similarly, the spectral function $\frac{1}{\pi} {\rm Im}
\Psi(t)$ on the r.h.s. of Eq.~\rf{eq:dispersio}
is also known in pQCD by analytical
continuation of $\Psi(q^2)$ to large
time-like values of $q^2=t >> \Lambda_{\rm QCD}^2$ up to order ${\cal O}(\alpha_s^4)$~\cite{CHETYRKIN},
\begin{equation}\label{chetim}
\frac{1}{\pi} {\rm Im} \Psi(t)|_{{\rm pQCD}}= 
\frac{N_c}{4\pi^2} \; \hat{m}^2(t) \; t \;
\left[ \; 1 + \frac{17}{3} \frac{\alpha_s(t)}{\pi} + \cdots\right]\,.
\end{equation}

In terms of physical degrees of freedom, the spectral function is given by the sum
\begin{equation}\label{espectral}
\frac{1}{\pi} {\rm Im} \Psi (t)= \sum_\Gamma 
| \langle 0 | S(0) |\Gamma \rangle|^2 \; (2\pi)^3 \delta^{(4)}\left( q- \sum p_{\Gamma}
\right)\,,
\end{equation}
where the sum over $\Gamma$ extends to all possible on--shell  states of the hadronic spectrum with the quantum numbers of the scalar current, including the  integration over their phase space.
These contributions add positively and, therefore, if the sum  is restricted to the lowest possible contribution of
two pions only, there will follow a rigorous lower bound to the full spectral function and hence, via the dispersion integral in Eq.~\rf{eq:dispersio} and the definition of the $S(x)$ current in  Eq.~\rf{eq:current}, to the quark masses. More precisely
\begin{equation}\lbl{eq:piopio}
\frac{1}{\pi} {\rm Im} \Psi(t) \geq
\frac{1}{\pi} {\rm Im} \Psi(t)|_{\Gamma=\pi \pi}=
\frac{3}{16\pi^2}
\sqrt{1- \frac{4 m_{\pi}^2}{t}} |F(t)|^2 \; \theta(t-4 m_\pi^2)\,,
\end{equation}
where $F(t)$ denotes  the $J=0$, $I=0$
scalar pion form factor defined by 
\begin{equation}
\langle \pi^a(p) \pi^b(p') | S(0) |0 \rangle =\delta^{ab} F(t)\,,\;\;\; 
t=(p+p')^2\,.
\end{equation}
Therefore, for sufficiently large values of $Q^2$ where pQCD in the evaluation of $\Psi''(Q^2)$ applies,
\begin{equation}\lbl{eq:basicin}
\frac{2}{3}N_c \frac{\hat{m}^2(Q^2)}{Q^2}
\left[ \; 1 + \frac{11}{3} \frac{\alpha_s(Q^2)}{\pi} + \cdots \right]\ge 	\int_{4m_{\pi}^2}^\infty dt \frac{1}{(t+Q^2)^3}
\sqrt{1- \frac{4 m_{\pi}^2}{t}} |F(t)|^2\,.
\end{equation}

As discussed in ref.~\cite{LdeRT97}, knowledge of the scalar pion form factor at the origin $F(0)$ and of its mean squared radius $\langle r^2\rangle_{s}^{\pi}$:
\be
F(t)=F(0)\left[1+\frac{1}{6} \langle r^2\rangle_{s}^{\pi}t+\cO(t^2)\right]\,,
\ee
results in a lower bound for $\hat{m}^2$ as a function of $Q^2$. A more restrictive bound was also derived in \cite{LdeRT97} using the fact that the phase of the scalar form factor $\delta_F(t)$ in the elastic region $4m_{\pi}^2 \le t \le 16m_{\pi}^2$ is precisely the $J=0$, $I=0$ $\pi- \pi$ phase--shift $\delta_{0}^{0}(t)$. Variations on similar analyticity properties concerning the strangeness changing scalar form factor have since then been also discussed in the literature (see e.g. ref.~\cite{CAPRINI} and references therein). 

The replacement of the full scalar spectral function by the r.h.s. of the inequality in Eq.~\rf{eq:piopio} is expected to be a gross underestimate  at large values of $t$. Indeed, for $t$ large,  pQCD predicts the spectral function to grow as  $\cO (\hat{m}^2 (t)\, t)$, while the scalar pion form factor  drops as $\cO (1/t)$ up to logarithms~\cite{BF73,LB80}. This suggests an improvement of the inequality in Eq.~\rf{eq:basicin} in the following way. First we separate the dispersion integral in the r.h.s. of Eq.~\rf{eq:dispersio} in two pieces
\begin{equation}\lbl{eq:disptwo}
\Psi''(Q^2)= \int_{(2 m_\pi)^2}^{t_0}  dt 
\frac{2}{(t+Q^2)^3} \; \frac{1}{\pi} {\rm Im}
\Psi(t)+\int_{t_0}^{\infty}  dt 
\frac{2}{(t+Q^2)^3} \; \frac{1}{\pi} {\rm Im}
\Psi(t)\,,	
\end{equation}
with the scale $t_0$ chosen sufficiently large so that the spectral function in the  high energy integral $t_0\leq t\leq\infty$ can well be approximated by its pQCD expression. Then we use the spectral function inequality \rf{eq:piopio} \underline{only} in the low energy integral $(2 m_\pi)^2\le t\le t_0$. This results in the following improved inequality

{\setl
\bea\lbl{eq:impineq}
\lefteqn{ \frac{2}{3}N_c \frac{\hat{m}^2(Q^2)}{Q^2}
\left[ \; 1 + \frac{11}{3} \frac{\alpha_s(Q^2)}{\pi} + \cdots \right]\ge  } \nn \\ & & \int_{4m_{\pi}^2}^{t_0} dt \frac{1}{(t+Q^2)^3}
\sqrt{1- \frac{4 m_{\pi}^2}{t}} |F(t)|^2   + \frac{4}{3}N_c \int_{t_0}^\infty dt \frac{\hat{m}^2(t)\; t }{(t+Q^2)^3}\; 
\left[ 1 + \frac{17}{3} \frac{\alpha_s(t)}{\pi} + \cdots\right]\,.
\eea}

\noi
We wish to emphasize that this inequality is rigorous, {\it provided that both $Q^2$ and $t_0$ are chosen sufficiently large}, so that pQCD can be applied to the evaluation of the function $\Psi''(Q^2)$ in the deep euclidean as well as to the dispersion integral in the $t_0\leq t\leq\infty$ interval. We next proceed to the evaluation of this bound.
 
\section{\normalsize Evaluation of the Improved Bound}\lbl{BSC}
\setcounter{equation}{0}
\def\theequation{\arabic{section}.\arabic{equation}}

The basic ingredient in the evaluation of the low--energy integral in Eq.~\rf{eq:impineq} is the Mushkhelishvili--Omnès representation of the form factor
\begin{equation}\lbl{eq:MO}
|F(t)|= |F(0)| \exp \left\{ \frac{t}{\pi} P\int_{4m_{\pi}^2}^{\infty} ds 
\frac{\delta_F(s)}{s (s- t)}\right\}\,,
\end{equation}
which follows from the analyticity properties of $F(t)$
if the form factor has no zeros and goes as $\cO(1/t)$ (up to logarithms) at large t. Then the phase $\delta_F(t)$ must approach $\pi$ asymptotically~\footnote{For a detailed discussion see e.g. ref.~\cite{PACO3}.}. The form factor at the origin is known from ChPT~\cite{GL84}
\be
F(0)=m_{\pi}^2 \left[1+\frac{m_{\pi}^2}{32\pi^2 F_{\pi}^2}(1-\bar{l}_3 )+\cO(m_{\pi}^4 ) \right]=m_{\pi}^2 (0.99\pm 0.02)\,,
\ee
where the error here takes into account a generous error on the low--energy constant $\bar{l}_3$ as well as on higher order corrections~\cite{Higor}.
\begin{figure}[h]

\begin{center}
\includegraphics[width=0.9\textwidth]{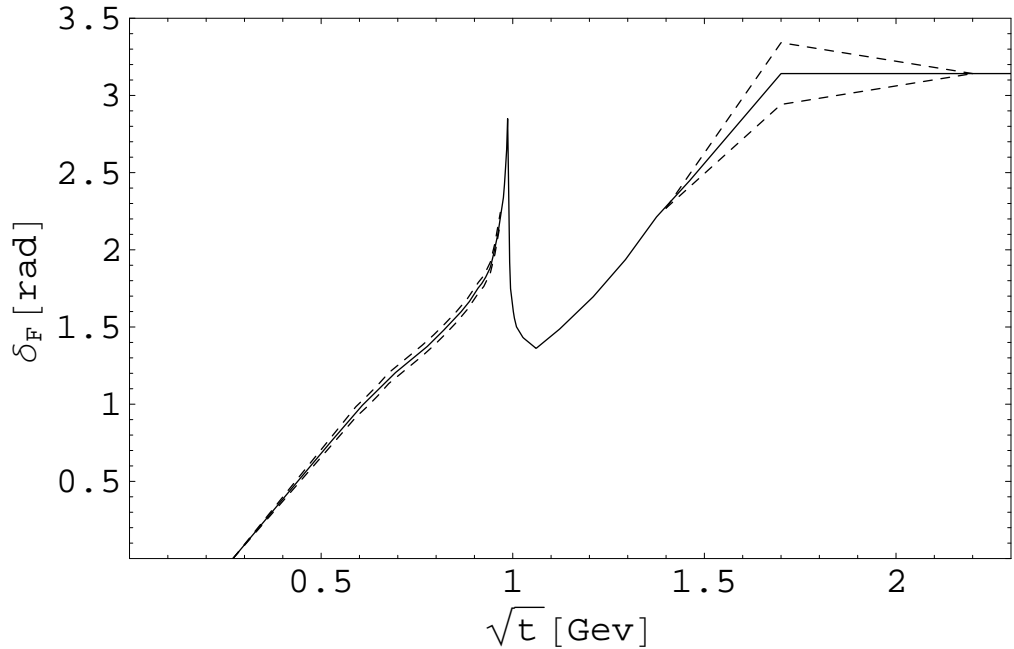}

\vspace*{0.5cm}
\caption {\it Shape of the phase $\delta_{F}(t)$ as a of function of $t$. In the interpolation region the error varies smoothly
up to 20\% at $\sqrt{t}=1.7~\MeV$. The errors below the $K-\bar{K}$ threshold
are of the same size as those given in ref.~\cite{PACO3}.}
\end{center}

\end{figure}
As already mentioned, the phase $\delta_{F}(t)$ in the region  $4m_{\pi}^2\leq t\leq 16 m_{\pi}^2$ is given by the $\pi \pi$ phase--shift $\delta_{0}^{0}(t)$  and it has been extracted from  $\pi-\pi$ scattering experiments~\cite{deltapipi}, from $K_{l4}$ decays~\cite{deltaK4} and $K\ra\pi\pi$ decays~\cite{deltaKpp}.  In fact the relation $\delta_{F}(t)=\delta_{0}^{0}(t)$ can be extended up to
the $K \bar{K}$ threshold, since inelastic processes do not play a
major role in the interval $16 m_{\pi}^2 \leq t \leq 4M_{K}^2$~\cite{ACCGL04}. Excellent  fits to the phase $\delta_F(t)$ in this region can be found e.g. in refs.~\cite{PACO,PACO2}. The opening of the  $K \bar{K}$ threshold, however, produces a square root singularity at $t=4M_{K}^2$ which causes a dip in the elasticity in the region $1~\GeV\le \sqrt{t}\le 1.1~\GeV$. The corresponding effect in the determination of the phase $\delta_{F}(t)$ which appears in Eq.~\rf{eq:MO} was first discussed in ref.~\cite{DGL90} within a two--channel representation, where only the $\pi\pi\ra K \bar{K}$ transition is assumed to drive the whole effect of the inelasticity. It was later confirmed by a more elaborate analysis in ref.~\cite{Mou00}, where a variety of parameterizations~\cite{param} are taken into account. As a result, the overall shape of the phase $\delta_F(t)$ is now rather well known up to energies $\sqrt{t}\lesssim 1.4~\GeV$. Beyond that, it can be assumed to join smoothly the asymptotic regime predicted by pQCD~\cite{BF73,LB80}. 
In Fig.~1 we show the plot of the phase $\delta_{F}(t)$ thus obtained (the solid line), where we also show a cautious  error margin of $\sim 20\%$ in the interpolation region: $1.4~\GeV\lesssim\sqrt{t}\lesssim 2.2~\GeV$ (the dotted lines). 
Fortunately,
this alleged source of uncertainty  turns out
to have little influence on the light quark mass bounds, as far as the interpolation is smooth.
This is because  the low--energy integrand in Eq.~\rf{eq:impineq} is modulated by three powers 
of $(t+Q^2)$ in the denominator which, because of the large $Q^2$--values we are considering,  makes the integral rather insensitive
to the precise values
of the phase in this intermediate region~\footnote{We wish to recall that the value of 
$\delta_F(t)$ in this intermediate region has been a source of debate in the literature~
\cite{PACO3,PACO4,PACO5,OR07},
where it is argued that beyond $\sqrt{t}\simeq 1.4~\GeV$ other
channels like the four pion channel may sensibly distort the shape of the
phase. 
The debate
focuses on the precision 
that the mean squared radius $\langle r^2\rangle_{s}^{\pi}$ can be extracted from the integral 
\be
\langle r^2\rangle_{s}^{\pi}=\frac{1}{|F(0)|}\frac{6}{\pi}\int_{4m_{\pi}^2}^\infty ds \frac{1}{s^2}\delta_{F}(s)\,.
\ee
We would like to stress the fact that, contrary to this observable,  the integral we are concerned with is much less sensitive to extrapolation uncertainties. In any case, the result we obtain  for the mean square radius, using the solid curve in our Fig.~1
is in good agreement with the value 
$\langle r^2\rangle_{s}^{\pi} = (0.61 \pm 0.04)~{\rm fm}^2$ quoted in \cite{ACCGL04}.}.
\begin{figure}[h]

\begin{center}
\includegraphics[width=0.9\textwidth]{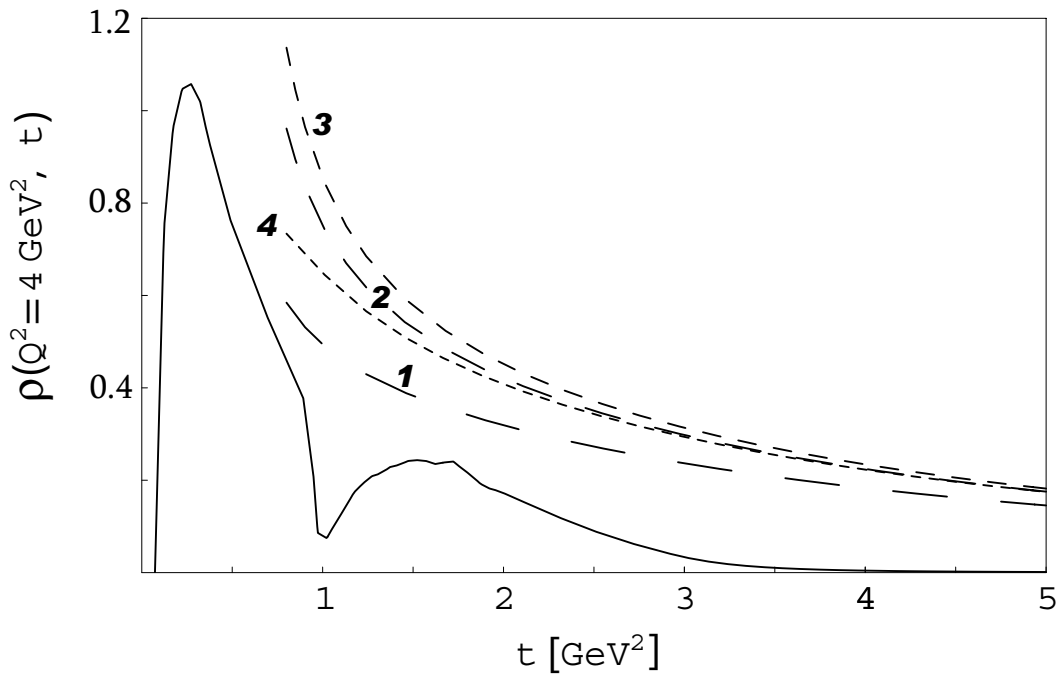}

\vspace*{0.5cm}
\caption {\it Shape of the low--energy integrand $\rho_{\rm low}\left(Q^2,t\right)$ (the solid curve) in Eq.~\rf{eq:low} and the high--energy integrand $\rho_{\rm high}\left(Q^2,t\right)$ (the dotted curves) in Eq.~\rf{eq:high} as a function of the $t$ for the choice $Q^2 = 4~\GeV^2$ . The four dotted curves correspond to the successive pQCD approximations.}
\end{center}

\end{figure}

The solid curve in Fig.~2 shows the shape (without errors)  of the low--energy integrand in Eq.~\rf{eq:impineq} 
\be\lbl{eq:low}
\rho_{\rm low}\left(Q^2,t\right)=\frac{1}{(t+Q^2)^3}
\sqrt{1- \frac{4 m_{\pi}^2}{t}} |F(t)|^2 \,,
\ee
as a function of $t$ for a reference choice $Q^2=4~\GeV^2$,
with the form factor $|F(t)|$ evaluated as described in the previous paragraph. The dashed curves in the same figure show the shape of the high--energy integrand: 

{\setl
\bea\lbl{eq:high}
\rho_{\rm high}\left(Q^2,t\right) & = & \frac{4}{3}N_c  \frac{\hat{m}^2(t)\; t }{(t+Q^2)^3}\; 
\left\{ 1 + \frac{17}{3} \frac{\alpha_s(t)}{\pi} + (35.94 -1.359\; n_f)\left(\frac{\alpha_s(t)}{\pi}\right)^2 + \right.\nn \\
& & (164.14 - 25.77 \; n_f + 0.259 \; n_f^2)\left(\frac{\alpha_s(t)}{\pi}\right)^3 +\nn \\
 & & \left. (39.34 -220.9 \; n_f +9.685 \; n_f^2 - 0.0205 \; n_f^3)\left(\frac{\alpha_s(t)}{\pi}\right)^4 +\cO\left[ \left(\frac{\alpha_s(t)}{\pi}\right)^5\right]\right\}\,,
\eea}

\noi
at the same reference choice $Q^2=4~\GeV^2$,
for different approximations of the pQCD series in powers of $\frac{\alpha_s(t)}{\pi}$; i.e., from one power (the curve with the largest dashing) to four powers (the curve with the shortest dashing). Each curve here is modulated by the appropriate running quark mass $\hat{m}^2(t)$ resulting from the inequality in Eq.~\rf{eq:impineq} and approximated at the corresponding number of loops.

\begin{figure}[h]

\begin{center}
\includegraphics[width=0.9\textwidth]{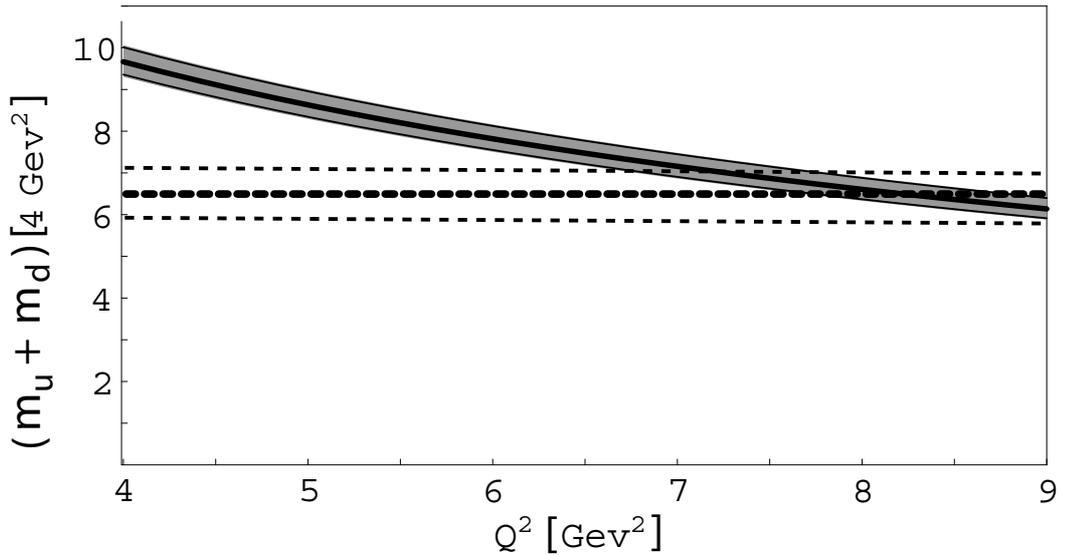}

\vspace*{0.5cm}
\caption {\it Plot of the lower bound (the gray band)  obtained for $[m_{u}+m_{d}](\mu=2~\GeV)$  in $\MeV$ as a function of $Q^2$ in $\GeV^2$, where we have also set $t_0 =Q^2$ in Eq.~\rf{eq:impineq}. The width of the band reflects the propagation of errors, mostly from the determination of the phase $\delta_{F} (t)$ shown in Fig.~1. The horizontal dashed band corresponds to the most recent lattice results quoted in the text.}
\end{center}

\end{figure}
We find that for $t\gtrsim 4~\GeV^2$ the error between the $\cO(\alpha_s^3)$ and the $\cO(\alpha_s^4)$ approximations to $\rho_{\rm high}\left(Q^2,t\right)$ is less than 5\%. It seems then natural to adopt the simplest choice $t_0 = Q^2$ for the separation scale $t_0$ in Eq.~\rf{eq:impineq} and proceed to the numerical evaluation of the bounds as a function of only one large scale. The corresponding lower bounds for $[m_{u}+m_{d}](\mu=2~\GeV)$ as a function  of $Q^2 =t_0$ in the interval: $4~\GeV^2 \le Q^2 \le 9~\GeV^2$  are then shown in Fig.~3. The thickness of the band indicates the effect of a scan in the propagation of the errors discussed above. They are largely dominated by the error in the extrapolation of the form factor. The choice $Q^2\ge 4~\GeV^2$ is already large enough for the error to be almost independent of $Q^2$. In other words, with the input discussed above, the bound at $Q^2=4~\GeV^2$ should already be a rigorous bound.

\section{\normalsize Conclusions}\lbl{con}
\setcounter{equation}{0}
\def\theequation{\arabic{section}.\arabic{equation}}

It is interesting to compare the values that we obtain for the lower bounds on $m_{u}+m_{d}$, as shown in Fig.~3, to the latest determinations of the light quark masses reported from lattice 
simulations:

{\setl
\bea
m_u +m_d & = & 6.6 \pm 0.6~\MeV \qquad {\rm MILC~Collaboration~\cite{MILC}} \\
m_u +m_d & = & 6.4 \pm 0.6~\MeV \qquad {\rm HPQCD~Collaboration~\cite{HPQCD}}\,.
\eea}

\noi 
These numbers correspond to masses in the $\overline{{\rm MS}}$--renormalization scheme at a reference scale $\mu=2~\GeV$, like our bounds in Fig.~3. The quoted  error, is our addition in quadrature of the statistical, lattice systematics, pQCD, and electromagnetic/isospin effects quoted in the original papers. These two lattice results are in remarkable agreement with each other. They correspond to the dashed band in our Fig.~3. As seen from this comparison, they are {\it somewhat smaller} than  the lower bounds. We find e.g.,

{\setl
\bea
m_u +m_d & \ge & 9.7 \pm 0.4~\MeV \qquad {\rm at} \qquad Q^2=4~\GeV^2\,, \\
m_u +m_d & \ge & 7.5 \pm 0.3~\MeV \qquad {\rm at} \qquad Q^2=6.5~\GeV^2\,, \\
m_u +m_d & \ge & 6.1 \pm 0.3~\MeV \qquad {\rm at} \qquad Q^2=9~\GeV^2\,.
\eea}

\noi 
One has to go to values as high as $Q^2 \gtrsim 7~\GeV^2$ to be in agreement. On the other hand, the bounds can very well accommodate some of the QCD sum rules results in the literature, like e.g.,

{\setl
\bea
m_u +m_d & = & 9.2 \pm 2.4~\MeV \qquad \cite{DN98} \\
m_u +m_d & = & 9.4 \pm 1.8~\MeV \qquad \cite{Pra98}\,.
\eea}

\noi
Lower values than these ones, however, have also been reported more recently, like e.g.,
\begin{equation}
m_u +m_d  =  7.5 \pm 0.7~\MeV \qquad \cite{JOP06}\,.	
\end{equation}
It will be interesting to see in which direction the future determinations, both from lattice QCD and from QCD sum rules,  will evolve.

\vspace*{1cm}

{\bf Acknowledgments}

\vspace*{0.25cm}

\noi
We are grateful to Laurent Lellouch for useful comments and for reading the manuscript.
The work of ADC and JT has been supported in part by the MEC Grant No. CYT FPA 2004-040582-C02-01, FIS 2004-05639-C02-01 as well as from the CIRIT Grant No. 2005 SGR-00564. The work of EdeR has been supported in part  by the European Community's Marie Curie Research Training Network program under contract No. MRTN-CT-2006-035482 (FLAVIAnet).

\end{document}